 \documentclass[final,3p,times,twocolumn]{elsarticle}
\usepackage{amssymb}
\usepackage{amsbsy}

\journal{Astroparticle Physics}

\begin{document}

\begin{frontmatter}


 \ead{maria.giller@kfd2.phys.uni.lodz.pl}

\title{An extended universality of electron distributions
      in cosmic ray showers of high energies and its application}


\author{Maria Giller, Andrzej \'Smia\l{}kowski and Grzegorz Wieczorek}

\address{The University of Lodz, Faculty of Physics and Applied Informatics, Pomorska 149/153, 90-236, Lodz, Poland}

\begin{abstract}
It is shown that the shape of any electron distribution in a high energy air shower is the same in all such showers, if taken at the same age, independently of the primary energy, mass and thus, of the interaction model. A universal behaviour has been also found within a single shower, such that the lateral distributions of electrons with fixed energies, at various shower ages, can be described by a single function of only one variable. The angular distributions of electrons with a fixed energy can be represented, at a given lateral distance, by a function of the product $\theta\cdot E^{\alpha}$ only, which is explained by a model of small angle electron scattering with  simplified energy losses. These results have been obtained by Monte Carlo simulations of the extensive air showers.
The electron universality can be used as a method for determining the longitudinal profile of any single shower from its optical images measured by the fluorescence light technique, which is particularly useful with showers observed with large fraction of the Cherenkov light.
\end{abstract}
\end{frontmatter}

\section{Introduction}
The subject of this paper is a study of various distributions of extensive air shower electrons, of both signs, at different levels of shower development. We concentrate on the highest energy end of the primary energy spectrum, at $E_0 \gtrsim 10^{16\div 17}$ eV. This energy region is of particular interest for several reasons. It is mainly the fact that the measured energy spectrum displays several structures, as compared to the straight power-law line in the log-log scale, such as the second knee, the ankle and the apparent GZK cut-off \cite{bib1}. Each of these structures may be, and probably is, an important imprint of the acceleration and/or propagation mechanisms of cosmic rays. In addition, the arrival directions of the primary particles, i.e. the shower axes, should indicate towards the particle sources, however, probably only for $E_0/Z > 10^{19}$ eV, where $Z$ is the primary particle charge.

The only way to study high energy cosmic rays is to observe the extensive air showers produced by them in the atmosphere. The most precise method for a determination of the primary energy $E_0$ on the event by event basis is to measure the optical light emitted by secondary particles of the shower while they propagate through the atmosphere. They excite the atmospheric nitrogen which in turn emits isotropically the fluorescence light, so that the air showers can be observed from the side at distances of several to some tens of kilometres. The fluorescence technique enables measurements of light images of a shower from which the longitudinal shower profile - the number of particles, or their energy deposit, as a function of depth in the atmosphere - can be deduced. This is possible because experiments show that the fluorescence light emitted by an air shower path element is proportional to the energy loss of shower particles along this element \cite{bib2}. Having determined the longitudinal shower profile one can calculate the energy of the primary particle by integrating the energy deposit along the shower track in the atmosphere in a way almost independent of its mass and the actual, although unknown, high energy interaction characteristics.

The fluorescence technique was used successfully for the first time by the Fly's Eye experiment \cite{bib3} and by its successors, the various set-ups of HiRes (see e.g. \cite{bib4}). More recently it has been applied in the Pierre Auger Observatory \cite{bib5,bib6} and the Telescope Array \cite{bib7}, together with large arrays of particle detectors.

A reconstruction of the primary energy $E_0$ of an air shower is, however, complicated by the fact that about one third of shower electrons also emit Cherenkov (Ch) light \cite{bib8}; the number of Ch photons produced by a high energy particle is 5$\div$6 times larger than its fluorescence photon yield.
Since the direction of the Ch emission is almost the same as that of the electron velocity, and the electron velocity is almost that of the shower, most Ch photons accumulate and propagate together with shower particles.
Those scattered in the atmosphere to the side by the Rayleigh and Mie processes constitute a non-negligible fraction of the total light registered by a telescope. Moreover, when an air shower is observed at an angle to its axis $\delta<30^{\circ}$, the flux of the so called direct Ch photons, i. e. those produced at the observed level, becomes also important \cite{bibnagwat}.
In addition, the photons from all the above processes are subject to scattering in the atmosphere on their way from the shower to the detector, causing blurring of the image. An analytical treatment of this effect \cite{bibmscatt}, as well as that based on simulations \cite{bibmcarlo}, have been worked out.

 Thus, the question arises how to reconstruct correctly the longitudinal shower profile i.e. the depth of the shower maximum $X_{max}$ - correlated with the mass of the primary particle - and its energy $E_0$.
The reconstruction method used by the Pierre Auger Collaboration \cite{bib9} is based on assuming as a first step an approximate relation between the Ch component and the energy deposit. The obtained shower profile is then subject to iterations to allow for a more accurate treatment of the Ch light. This method works very well for the air showers registered by the hybrid detector i.e. by at least one telescope of the Fluorescence Detector and one station of the Surface Detector, an array of 1660 water-Cherenkov tanks. However, for this  approach to work correctly the fraction of the Ch contribution in the telescope should not exceed (50 -70) $\%$ of the fluorescence light.

The method for the reconstruction of the longitudinal shower profiles proposed in this paper is based on the universality of $electron$ distributions in the air showers. 
By the universality of electron distributions we mean two things:
\begin{enumerate}
\item the distributions of electrons within one shower can be described by less than five parameters, being combinations of the five variables:
shower age $s$, electron energy $E$, lateral distance $r$ and the two component angle $\vec\theta$ to the shower axis;
\item the shape of any electron distribution is the same in \emph{any} high-energy shower if taken at the same shower age $s$, independently of the primary energy and mass.
\end{enumerate} 

The word “universality” was also used with different meanings, e.g. concerning the muon component \cite{bib10,bib11} or a particular detector of shower particles, as in e.g. \cite{bib12}. Here, however, we are concerned with shower electrons only. 

In this paper we show that the electron distributions in the high energy showers are indeed universal in both above meanings.
To prove this we study, by full simulations of proton and iron initiated showers with $E_0=10^{16}$ and $10^{17}$ eV, the electron lateral distributions with fixed energies and their angular distributions as a function of the lateral distance. 
We find that the lateral distributions of electrons with different energies at various shower ages can be represented with a good accuracy by a \emph{single function}, i.e. a universal one in the first sense, of a variable proportional to $r\cdot E^{0.53}\cdot e^{-0.893(s-1)}$, where $r$ is the electron lateral distance in $g\,cm^{-2}$. 
In addition, the angular distributions of electrons with energy $E$ at a given distance $r$ are functions of  $r$ and $\vec\theta\cdot E^{0.73}$ only.
These new universal characteristics, determined in this paper, constitute an extension of the meaning of the electron universality as described in point 1. We show that these distributions are the same for any simulated shower, so that they are universal in the second sense for $E_0=10^{16}\div 10^{17}$ eV.

Next, we have extended the energy range of this universality up to $E_0=10^{20}$ eV by comparing the lateral distributions with earlier studies \cite{bib20,bib21,bib22} based on Monte-Carlo shower simulations with the thinning procedure.

Universality can be applied to the reconstruction of showers without any restriction on the Cherenkov contribution to the total light. It enables an exact prediction of a shower image on the telescope camera at any time interval, once the shower development state, i.e. its age $s$ and the total number of electrons  $N_e$ at this time are known.
We show that to correctly reconstruct the longitudinal profile of an air shower with \emph{any} geometry from its optical images, in principle all the electron distributions have to be known. This work simplifies their description and shows that our predictions can be applied to any high-energy shower.

We also give formulae connecting the lateral distance and the angle of an emitting electron in the air shower with the position of the photons on the camera of an imaging telescope.

\section{Universality of the electron distributions in the extensive air showers}
\subsection{The universality so far}
Hillas \cite{bib14} suggested that by analogy to the pure electromagnetic cascade
an age $s$ of the hadronic shower at a slant atmospheric depth $X$ in $g\,cm^{-2}$, defined as
\begin{equation}
s= 3X/(X+ 2X_{max})
\end{equation}
might be useful for a description of shower particles. The advent of fast computers made it possible to use Monte-Carlo methods for simulations of individual hadronic showers and to obtain all the relevant particle distributions at different ages. By using the shower simulation tool, CORSIKA \cite{bib15}, it was shown \cite{bib8} that the shapes of the energy spectra of electrons on a given level of shower development depend on the age $s$ of this level, as it is in the purely electromagnetic cascades. They do not depend on the energy or mass of the primary particle. (Of course, the highest energy end of the spectrum does depend on the primary energy $E_0$, but the independence concerns the bulk of electrons, i.e. those with energies around the critical energy of the air, $\sim$ 80 MeV, up to, say, $\sim$1 GeV).

It is important to note that if any shower characteristic is independent of the primary particle mass it must be also independent of the interaction model adopted. Indeed, it can be imagined that an iron shower is actually initiated by a proton with drastically changed interaction model: the first interaction cross-section is much larger - such as that for an iron nucleus, the dissipation of the primary energy is much quicker - as it is in an iron shower, and so on. Thus, if a shower characteristic is unchanged despite such drastic changes in the proton interaction model it will be even less affected by the differences between various models of the proton interaction.

Moreover, what is important for our purposes here, it was shown in [8] that for big showers, where the number of electrons in a given energy interval is large, the electron energy spectrum fluctuates very little from shower to shower.

Other universal characteristic, determined first in \cite{bib17}, is that the angular distribution of shower electrons \emph{with a given energy} at a particular age $s$ depends of this energy only. It does not depend even on the age $s$. 
This universal feature simplified a calculation of the angular distribution of the Cherenkov light emitted by shower electrons as a function of $s$ and height in the atmosphere, done by the integration of such a distribution for a given $E$ subject to weighting by the energy spectrum and the Ch yield at this level \cite{bib18}.

Yet another universal feature of electrons in the air showers is their lateral distribution at a given level $s$. When electrons with all energies are considered, the shape of their lateral distributions, with the lateral distance taken in Moli\`ere units, depends on $s$ only \cite{bib19}. The lateral distributions of electrons with fixed energies were studied first in \cite{bib20} showing that also they depend on $s$ only (Fig. 1).

The idea of the shower universality was explored by other authors as well. Nerling et al.\cite{bib21} fitted the electron energy distributions as a function of $s$ by a relatively simple function. They studied and parametrised the angular distribution of electrons and that of the produced Ch light as functions of height in the atmosphere and $s$.

In a later study Lafebre et al.\cite{bib22} analysed further the angular and lateral distributions for fixed electron energies, providing another parametrisation (but see 2.2.1).
They also analysed fluctuations from shower to shower of the distributions of electron energy, angle and lateral distance. They found that for the bulk of electrons they were very small, confirming their universal nature. Thus, the electron universality has been used so far in the second sense i.e. as an independence of the primary particle parameters.

As it will be shown below all the above described distributions are not sufficient for our final purpose which is to predict, as accurately as possible, optical shower images.
\begin{figure}[t]
  \centering
  \includegraphics[width=0.50\textwidth]{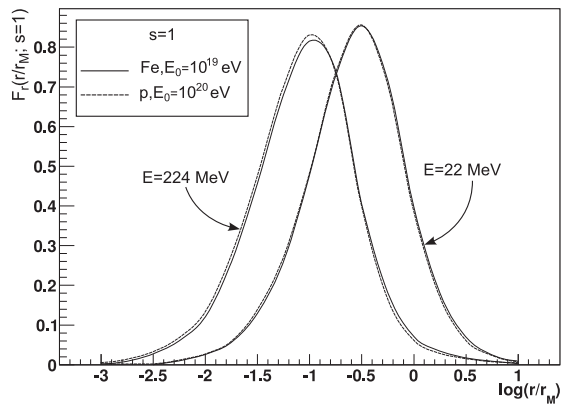}
  \caption{ Independence of the lateral distributions of electrons with two fixed energies $E$ of the primary particle \cite{bib20}. Shower maximum. Each curve is an average of 20 showers. }
  \label{fig1}
 \end{figure}
\begin{figure}[t]
  \centering
  \includegraphics[width=0.50\textwidth]{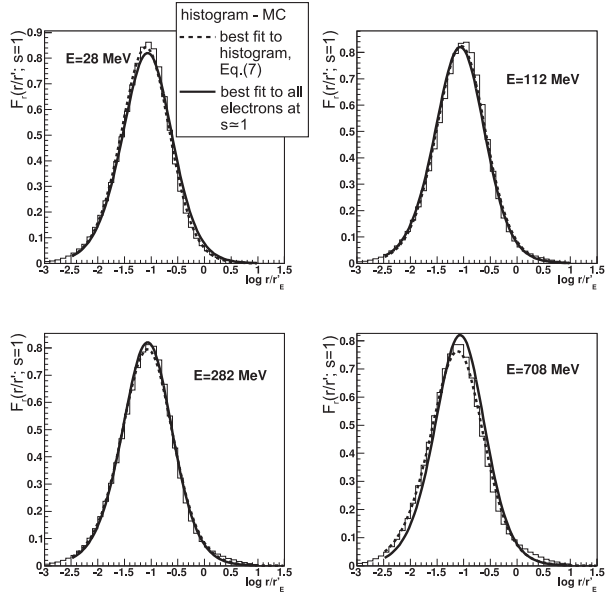}
  \caption{ Independence of lateral distributions $F_r(r/r'_{E};s\simeq 1)$ of electron energy $E$, for four values of $E$. One $10^{17}$ eV iron shower. }
  \label{fig2 }
\end{figure}
\subsection{Extending the air shower universality}
We start with showing that the universality in the first sense holds for one $10^{17}$ eV iron shower. Then, we will demonstrate that it holds also in the second sense.

Thus, our first aim is to describe, in a possibly \emph{simple} way, the function $f(\vec\theta ,r,E,s)$, a 5-dimensional distribution of electrons. We can represent this function as follows
\begin{equation}
	f(\vec\theta ,r,E,s)= f_E(E;s)\cdot f_r(r; E,s)\cdot f_{\theta}(\vec\theta; r,E,s)\textnormal{.}
\end{equation}
The variables on the r.h.s. of the semicolons are just the parameters of the functional dependence on the variable before the semicolon.
The normalisations of the above functions are such that
\begin{eqnarray}
	\int_0^{E_0} f_E(E;s)\,dE=1 \;\textnormal{,}      \nonumber \\
	\int_0^{\infty} f_r(r;E,s)\,dr=1 \;\textnormal{,}      \nonumber \\
	\int_{4\pi} f_{\theta}(\vec\theta;r,E,s)\,d\Omega=1 \;\textnormal{.}
\end{eqnarray}
Thus, the normalisation of the function $f$ at any level $s$ is
\begin{equation}
\int\int\int f(\vec\theta,r,E,s)\,d\Omega dr dE=1
\end{equation}
and the number of electrons $dN_e$ on level $s$ with energies $(E,E+dE)$ , at the lateral distance $(r,r+dr)$ and with angles $\vec\theta$ within $d\Omega$ equals
\begin{eqnarray}
	dN_e=N_e \,f_E(E;s)\cdot f_r(r;E,s)\cdot \nonumber \\
\cdot f_{\theta}(\vec\theta,r,E,s) \;d\Omega dr dE \;\textnormal{,}
\end{eqnarray}
where $N_e$ is the total number of electron at age $s$.
It is only $f_{\theta}$ which is still unknown. However, we start with studying $f_r(r;E,s)$ again since we have found an interesting description of it.
\subsubsection{Lateral distributions of electrons with fixed energies}
As the lateral distribution (LD) of electrons with all energies, as well as the energy distribution $f_E (E; s)$ depend on $s$ only it is expected that the only parameters on which the LD of electrons with a fixed energy $E$ should depend are  $E$ and $s$.
Indeed, it is the case, as it was shown in \cite{bib20} by using shower simulations with CORSIKA.
The distance scale, $r_E$, was chosen to be inversely proportional to $E$. 

Now we propose a better scale
\begin{equation}
	r'_E=X_0\, \left( \frac{21MeV}{E} \right)^{\alpha} \textrm{ with } \alpha<1 \;\textnormal{,}
\end{equation}
where $X_0$ is the radiation length of the air.

First we study LD of electrons with fixed $E$ around the shower maximum. 
Analysing full simulation results of one iron shower with $E_0=10^{17}$ eV we have found that if the LD of electrons with a fixed $E$ is presented as a function of $r/r'_E$, with $\alpha=0.53$, it is almost independent of $E$. Fig. 2 illustrates this situation for the shower level $s\simeq 1$ (the actual level studied is $s=0.95$, but to stress that the level is very close to shower maximum we will often quote it as $s\simeq 1$). Each of the four graphs refers to a different fixed electron energy $E$ from the most populated regions.
The histogram represents the true distribution obtained from the shower simulation, and the solid line is the distribution of $r/r'_E$ i.e. that of $r\cdot E^{0.53}$ for all electrons on this level, so it is the same for all four graphs. The index $\alpha=0.53$ was chosen to best fit the distributions of $r\cdot E^{\alpha}$ for various $E$. It can be seen that the curve for all electrons (solid) describes well the histograms for particular values of $E$. Thus, instead of $f_r(r; E, s\simeq 1)$ describing many distributions for many energies, we have a single function $F_r(r/r'_E;s\simeq1)$ determined for all electrons on this level and normalized to unity afterwards.

We fit it by an analytical function of the Nishimura-Kamata-Greisen type, as it was done in \cite{bib20}:
\begin{figure}[t]
  \centering
  \includegraphics[width=0.50\textwidth]{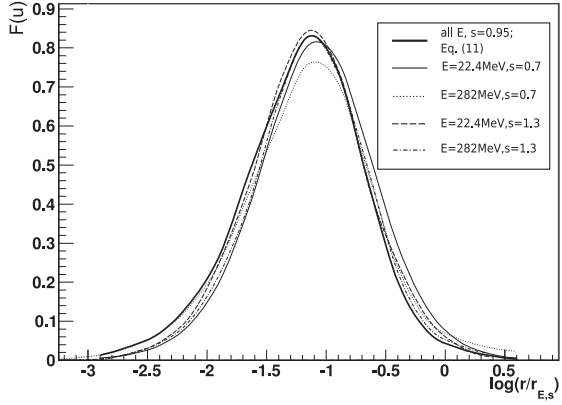}
  \caption{Comparison of the universal distribution $F(u)$ with parameters fitted to all electrons at $s=0.95$  with distributions of $u=r/r_{E,s}$ at $s\ne 1$. Values of $E$ and $s$ chosen from edges of their distributions. $F(u)$ determined from one $10^{17}$ eV iron shower. Other curves - averages from 20 showers each, initiated by $10^{19}$ and $10^{20}$ eV proton and iron nuclei \cite{bib20}.}
  \label{fig3}
\end{figure}

\begin{figure}[t]
  \centering
  \includegraphics[width=0.50\textwidth]{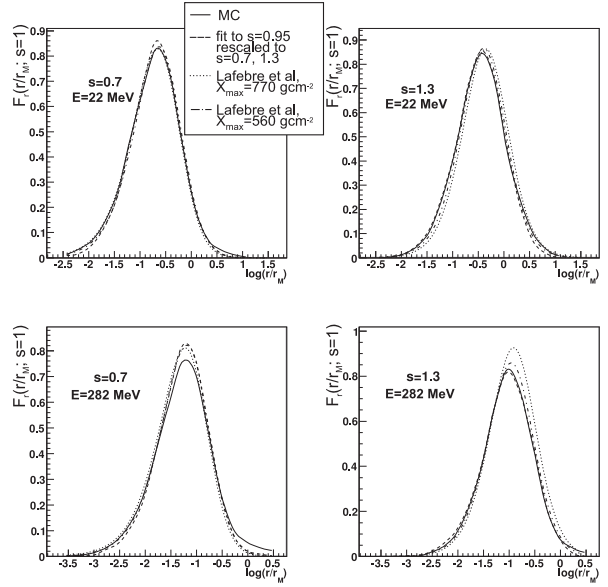}
  \caption{Comparison of the distributions of $r/r_M$ obtained by three ways for two electron energies at two ages. Lafebre fits done to averages of 20 showers.}
  \label{fig4}
\end{figure}
\begin{eqnarray}
    f_r(r;E, s=1) \,d\,r= \\ \nonumber
    F_r(r/r'_E; s=1) \,d\,\log\,r/r'_E= \\
=C\,\left(\frac{r}{r'_E}\right)^{\alpha}\left(1+k\frac{r}{r'_E}
\right)^{-\beta}\cdot d\,\log\frac{r}{r'_E} \;\textnormal{,}\nonumber
\end{eqnarray}

where $C=\ln(10)\cdot k^{\alpha} \frac{\Gamma(\beta)}{\Gamma(\alpha)\cdot \Gamma(\beta-\alpha)}$, and the parameters for $s=0.95$ are $k=10.0,\quad\alpha=1.694,\quad \beta=3.675$.

Since the fit differs a bit from the actual distributions for different $E$, we have also calculated the best fitting parameters $ \alpha$ and $\beta$ as functions of $E$ and approximated them as polynomials of $\log\,E(GeV)$. The corresponding coefficients are given in Table 1 and the dotted curves in Fig. 2 refer to these values.

\begin{table}[h]
\begin{center}
\caption{Dependence of parameters $\alpha$ and $\beta$ on $x=\log\,E(GeV)$ as $p=a\,x^3+b\,x^2+c\,x+d$; $k=10$.}
\begin{tabular}{c|c|c|c|c}
\hline coeff $\to $ & $a$ & $b$ & $c$ & $d$ \\
$\downarrow$parameter &  &  &  &   \\ \hline
$\alpha$ & 0.117 & 0.016 & -0.5 & 1.35\\ \hline
$\beta$ & 0.090 & 0.110 & -0.5 & 3.20\\ \hline
\end{tabular}
\label{table_1}
\end{center}
\end{table}

Next, we move to levels away from the shower maximum. 
In \cite{bib20} it was shown that the LD at a level $s \neq 1$ is almost the same as that for $s=1$ but shifted in the $\log\,r/r_E$ scale by $\Delta(s)$, where
\begin{equation}
	\Delta(s)=0.388\,(s-1)
\end{equation}
was independent of $E$.
Such a shift is equivalent to the rescaling of the $r'_E$ by introducing a new scale $r_{E,s}$
that includes different levels $s$
\begin{eqnarray}
           &&r_{E, s}=r'_E\cdot 10^{0.388(s-1)}= \\
		   &&=X_0\,\left(\frac{21MeV}{E} \right)^{0.53}\cdot e^{0.893(s-1)} \;\textnormal{.} \nonumber
\end{eqnarray}
Thus, we find that LD of electrons with any energy $E$ at the level $s$ can be described, in good approximation, by a single function $F(u)$ where
\begin{equation}
           u=r/r_{E,s}
\end{equation}
and
\begin{eqnarray}
    f_r(r;E,s)\,dr  = F(u)\,d\,\log\, u\simeq \\
\simeq 5.1\cdot 10^2\,u^{1.694}(1+10\,u)^{-3.675}d\,\log\, u \nonumber
\end{eqnarray}
with the same values of the parameters as those for $F_r$, Eq. (7).

Fig. 3 shows a comparison of $F(u)$ (thick, solid line) with the simulation results for two electron energies, $E\simeq 22.4$ MeV and $282$ MeV from $\Delta\,\log\,E=0.1 $, at two ages $s=0.7$ and 1.3, therefore far from shower maximum, and at both ends of the energy distribution, taken from \cite{bib20}. It is seen that even at these rather extreme values of $E$ and $s$ the universal curve describes the actual distribution approximately well. A better description is achieved by allowing for the energy dependence of the parameters $\alpha$ and $\beta$ (Tab. 1). Also a better agreement of the curves is achieved for $|s-1|\le 0.2$ and energies (40$\div$60) MeV, where there are most of shower electrons.

It should be noted that in Fig. 3 we compare average results\footnote{The aim of paper \cite{bib20} was not to study fluctuations from shower to shower. Anyway, since the primary energy used was ultra-high it was necessary to apply thinning in shower simulations which causes an increase of the electron distributions fluctuations. Thus, several showers were simulated for each  primaries.} for $E_0=10^{19}$ and $10^{20}$ eV proton and iron showers from \cite{bib20}, with our single $10^{17}$eV iron shower. The agreement confirms the universal character of $F(u)$ in the second sense.

Lafebre et al. \cite{bib22} also studied electron distributions in the air showers. They parametrised the distribution of $r/r_M$ for fixed $E$ as functions of the \emph{absolute} distance $t=(X-X_{max})/X_0$ from the shower maximum. Thus, to compare their results with ours we have to adopt some value of $X_{max}$. We choose two values: $X_{max}=770\,g\,cm^{-2}$ - referring to proton showers with $E_0\simeq 10^{19}$ eV, and $X_{max}=580\,g \,cm^{-2}$ - to iron with $10^{17}$ eV.
Their parametrised distributions for the two values of $X_{max}$ are shown in Fig. 4 together with the simulations (solid lines, from \cite{bib20}) and our best fitted curves (Eq. 7) rescaled to $s\ne 1$.
It can be seen that their description is slightly worse for $s=1.3$ and $E=282$ MeV. This may be caused by their using the variable $t$ to describe the universality, i.e. the independence of $E_0$. We claim that the universal distributions are functions of $s$, thus, of $(X-X_{max})/X_{max}$, not those of $X-X_{max}$. Certainly, at a level several hundreds $g \,cm^{-2}$ away from the maximum of a $10^{20}$ eV shower the electron distributions are closer to those at the maximum than the distributions at the same distance in $g \,cm^{-2}$ in a $10^{16}$ eV shower.

Finding $F(u)$ (Eq. 11) which describes the lateral distributions of the bulk of electrons with different energies at various ages within one shower we have confirmed the universality in the first sense, defined in the Introduction. But at the same time we have again confirmed the universality of $F_r(r/r'_E;s\simeq1)$ in the second sense, i.e. its independence of $E_0$ and mass. Indeed, in Fig. 4 Lafebre proton showers have $E_0=10^{18}$ eV and ours is an iron one with $E_0=10^{17}$ eV rescaled to $s\ne 1$ based on simulations for $E_0=10^{19}$ and $10^{20}$ eV proton and iron showers \cite{bib20}.
\begin{figure}[t]
  \centering
  \includegraphics[width=0.50\textwidth]{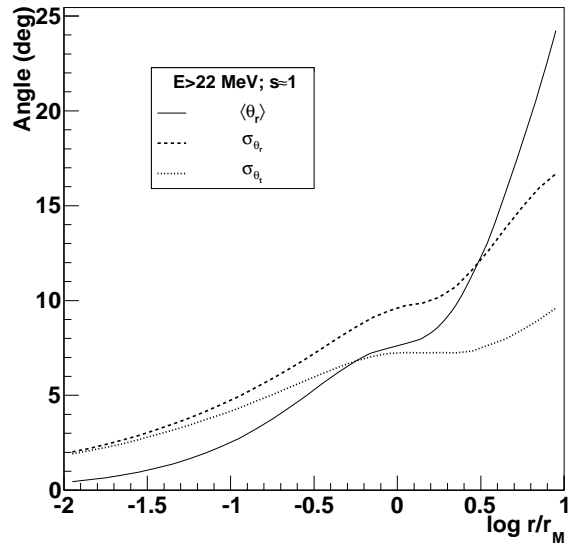}
  \caption{Dependence of the mean electron angle $\langle\theta_r\rangle$, its dispersion $\sigma_{\theta_r}$ and the dispersion $\sigma_{\theta_t}$ of $\theta_t$ on the radial distance $r/r_M$. }
  \label{fig5}
\end{figure}

\begin{figure}[t]
  \centering
  \includegraphics[width=0.50\textwidth]{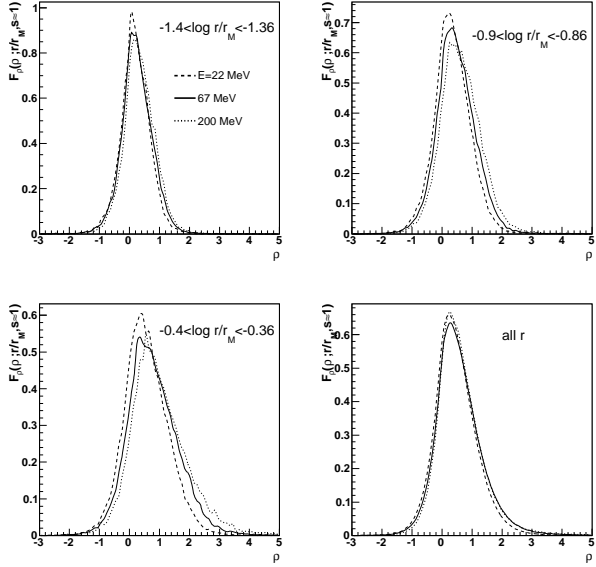}
  \caption{Independence of the distributions $F_{\rho}$ of $\rho=\theta_r\cdot E^{0.73}$ ($\theta_r $ in degrees, $E$ in GeV) of electron energy $E$; $s\simeq 1$.}
  \label{fig7}
\end{figure}
\subsubsection{Angular distributions of electrons with fixed energies at various lateral distances}
We present results concerning levels close to the shower maximum, i.e. $s\simeq 1$.

To study $f_{\theta} (\theta; r,E,s) $ we have used the fully simulated iron shower with  $E_0 = 10^{17} $eV. The number of electrons in the 4-dimentional intervals $\Delta \vec\Omega\cdot \Delta r\cdot E$ is much smaller than in the previously discussed distributions. Thus a full shower simulation was necessary to avoid the artificial fluctuations introduced by thinning.
The highest primary energy of an air shower to be fully simulated by us in a reasonable time is $10^{17}$ eV.

The angular distribution at some distance $r$ from the shower axis is not symmetric around it. For its description we have chosen the two angles: $\theta_r$, called radial angle, being the projection of the particle angle $\vec\theta$ on the plane containing the shower axis and vector $\vec r$, and $\theta_t$ - the tangential angle being the projection of $\vec\theta$ on the plane perpendicular to $\vec r$.

\begin{figure}[t]
  \includegraphics[width=0.50\textwidth]{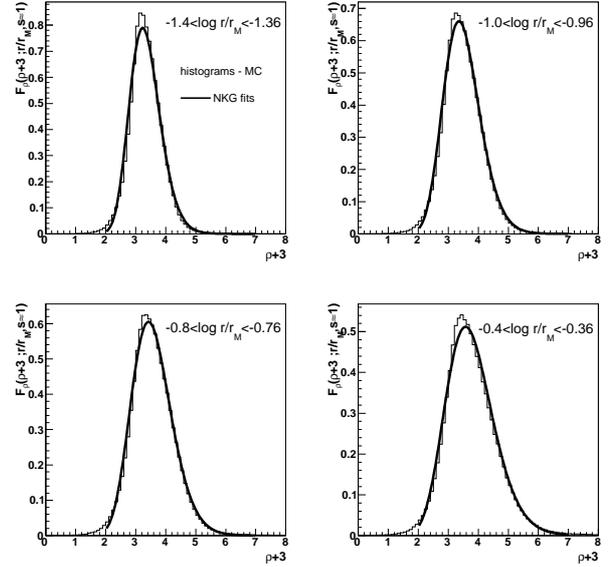}
  \caption{Fitting the simulated distributions $F_{\rho}(\rho +3; r/r_M, s\simeq 1)$ with Nishimura-Kamata-Greisen functions for four distance intervals.}
  \label{fig8}
\end{figure}
A clear demonstration of a dependence of the electron angular distribution on the lateral distance is shown in Fig. 5, where the mean angle
 $\langle\theta_r\rangle$ is presented as a function of the lateral distance $r/r_M$ in the Moli\`ere units for electrons with energies $E >$ 22 MeV ($\langle\theta_t\rangle\, = 0$). Also shown are the dispersions of both angles, increasing with the distance as well.

Next, we notice that the distributions of $\theta_r$ for different $E$ and any $r$ scale in such a way that the distribution of $\rho = \theta_r \cdot E^{\alpha}$ is almost independent of $E$. For $s$ = 0.95 we obtain $\alpha = $ 0.73. This is illustrated in Fig. 6, where the distributions of $\rho = \theta_r\cdot E^{0.73}$ are presented for three electron energies, from the region, where there are most of them, for some radial distances $r/r_M$ and for all $r$. (However, $\alpha$ seems to decrease slightly with $r$, but at this stage we adopt it as independent of $r$.)

To parametrise the distributions of \mbox{$\theta_r\cdot E^{0.73}$} for different $r/r_M$ a suitable function seems again that of the NKG form
\begin{eqnarray}
&&  F_{\rho}\left(x;\frac{r}{r_M},s\simeq 1\right)\,dx = \\
&&=\beta^{\mu}/B(\mu, \nu-\mu)\cdot x^{\mu-1}(1+\beta x)^{-\nu}\,dx \;\textnormal{,}\nonumber
\end{eqnarray}
where $B(\mu, \nu - \mu)$ is the Euler beta function. Since $\theta_r$ and $\rho$ may be negative we add to the variable $\rho$ a value 3 $deg\cdot GeV^{0.73}$ and fit the distributions of the above form with $x = \rho + 3$. The results are illustrated in Fig. 7, where the actual distributions of $x = \theta_r·E^{0.73}+3$ are presented for electrons at four distance intervals, together with the fitted
curves. We see that the fits describe well the histograms. (The fits with a gamma function were worse). The integrals over $x$ of all distributions are equal to unity. The dependence of $\mu$  and $\beta$ on \mbox{$y = \log(r/r_M)$} have been parametrised as 3-degree polynomials: \mbox{$a·y^3+ b·y^2+ c·y + d$}, and the coefficient values are given in Table 2. Value of $\nu$ has been set constant: $\nu=171$.

\begin{table}[h]
\begin{center}
\caption{Dependence of the parameters $\mu$ and $\beta$ (Eq. 12) and $\sigma_1$, $\sigma_2$ and $p$ (Eq. 14) as polynomials $ay^3+by^2+cy+d$, where $y=\log\,r/r_M$ $\nu =171$.}
\begin{tabular}{c|c|c|c|c}
\hline coeff $\to $ & $a$ & $b$ & $c$ & $d$ \\
$\downarrow$parameter &  &  &  &   \\ \hline
$\mu$  & - & 26.3 & 1.4 & 21.4\\ \hline
$\beta$ & -0.085 & -0.036 & -0.017 & 0.041 \\ \hline
$\sigma_1$ & - & - & 0.138 & 0.905 \\ \hline
$\sigma_2$ & - & -0.092 & -0.07 & 0.443 \\ \hline
$p$ & - & - & 0.168 & 0.463 \\ \hline
\end{tabular}
\label{table_2}
\end{center}
\end{table}

\begin{figure}[t]
  \centering
  \includegraphics[width=0.50\textwidth]{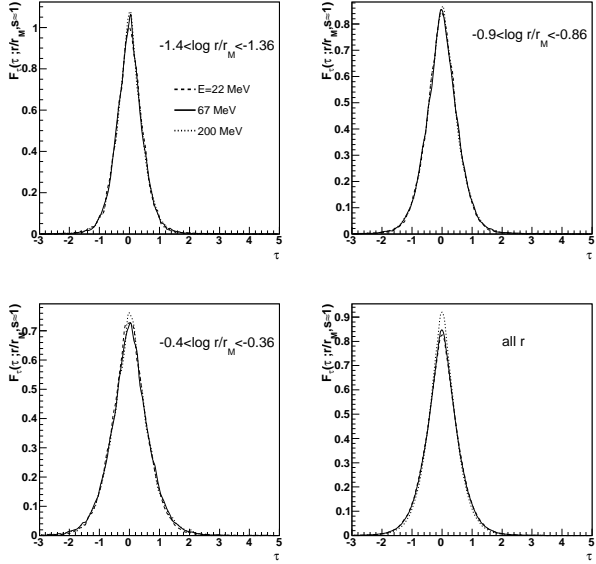}
  \caption{Independence of the distributions  $F_{\tau}$ of $\tau=\theta_t\cdot E^{0.73}$ of electron energy $E$; $s\simeq 1$.}
  \label{fig9}
\end{figure}

\begin{figure}[t]
  \centering
  \includegraphics[width=0.50\textwidth]{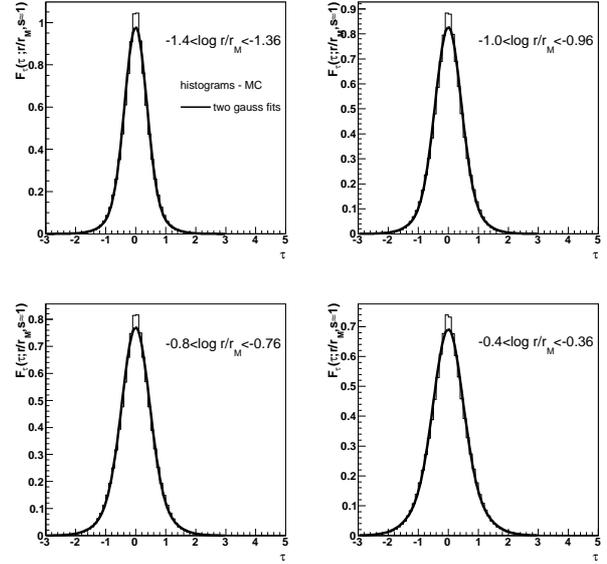}
  \caption{Fitting the simulated distributions $F_{\tau}(\tau;r/r_M, s\simeq 1)$ with a sum of two Gaussians for four distance intervals. }
  \label{fig10}
\end{figure}

\begin{figure}[t]
  \centering
  \includegraphics[width=0.50\textwidth]{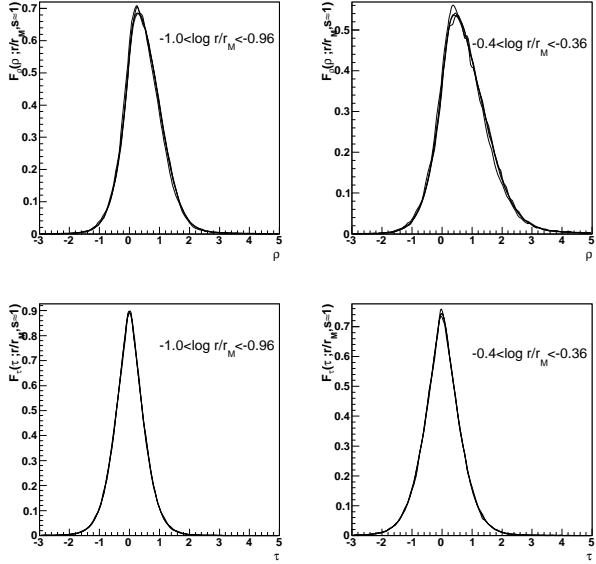}
  \caption{ Independence of $F_{\rho}(\rho;r/r_M,s\simeq 1)$ - two upper graphs for two distance intervals and $F_{\tau}(\tau;r/r_m,s\simeq 1)$ - two lower graphs - of primary energy and mass. On each graph there are 4  curves, each refering to proton or iron shower with $E_0=10^{16}$ or $10^{17}$ eV. }
  \label{fig0}
\end{figure}

\begin{figure}[t]
  \centering
  \includegraphics[width=0.5\textwidth]{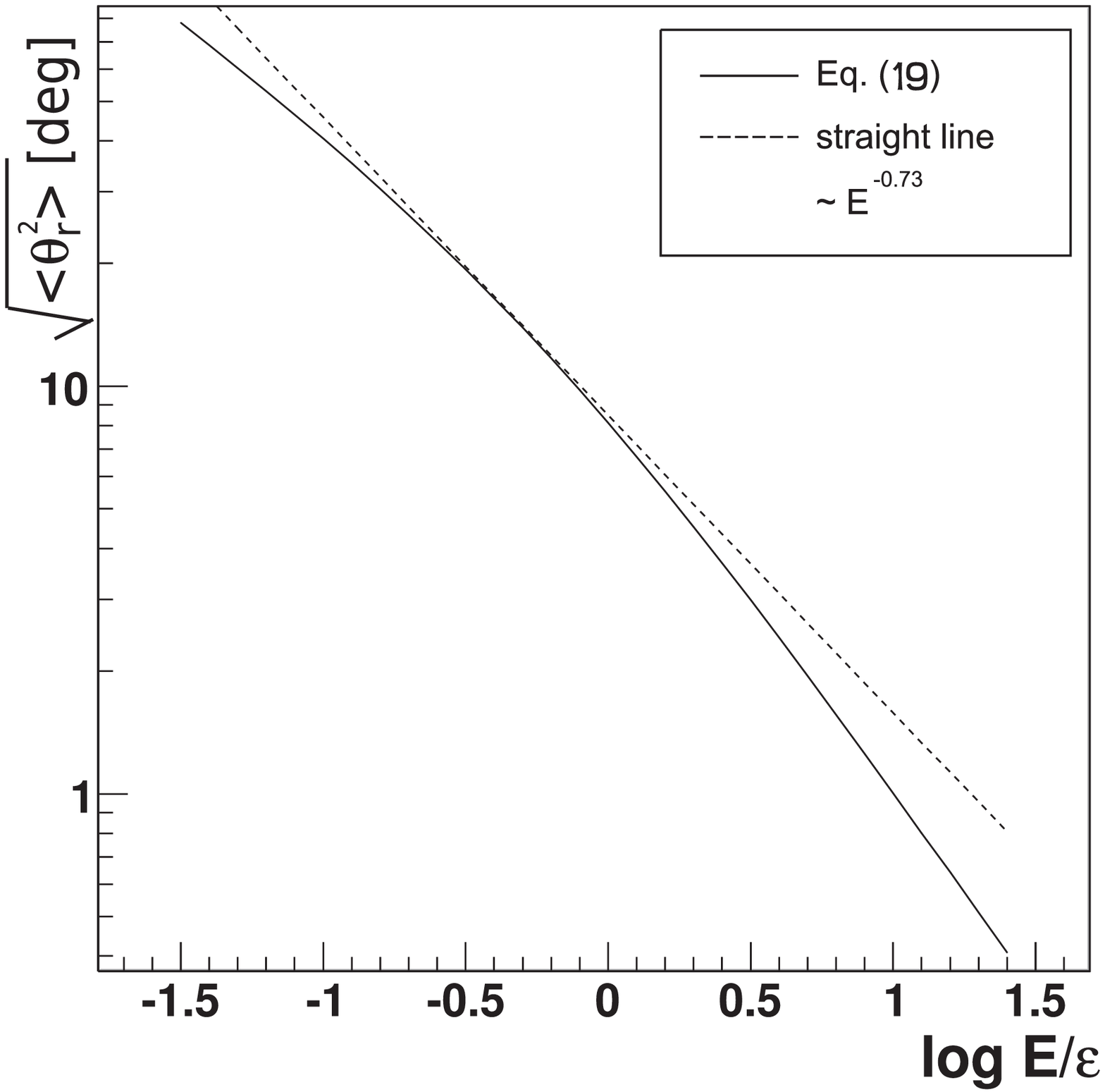}
  \caption{ Mean square radial angle $\sqrt{\langle\theta_r^2\rangle}$ as function of electron energy $E$ in units of the critical energy $\varepsilon$. }
  \label{fig6}
\end{figure}
Next, it is necessary to determine the distributions of the tangential angle $\theta_{t}$. At this stage of the analysis we assume that there is no correlation between $\theta_r$ and $\theta_t$ at a particular distance interval. Indeed, our study shows that the angular distributions, at a fixed $r$ and $E$, depend roughly on
\begin{equation}
	\eta=\sqrt{(\theta_r-\langle\theta_r\rangle )^2+\theta_t^2} \;\textnormal{.}
\end{equation}

If the distribution of $\eta$ per unit solid angle had been Gaussian, which is only approximately true, then the distributions of $\theta_r$ and $\theta_t$ would have been independent.

We first check whether the distributions of \mbox{$\tau = \theta_t\cdot E^{0.73}$} are independent of $E$ for each interval of $\log(r/r_M)$. This is found to be the case and the independence is even better fulfilled than that for \mbox{$\rho = \theta_r\cdot E^{0.73}$}. In Fig. 8 we present the results for three values of $E$, as in Fig. 6, for three distance intervals and for all $r$. 

We fit the distributions of $\tau$ for each distance interval with a sum of two Gaussian functions:
\begin{eqnarray}
&&F_{\tau}\left(\tau; \frac{r}{r_M},s\simeq 1\right)= \\
&&=\frac{1}{\sqrt{2\pi}}\left[\frac{p}{\sigma_1}\cdot exp\left(\frac{-\tau^2}{2\sigma_1^2}\right)+\frac{1-p}{\sigma_2}\cdot exp\left(\frac{-\tau^2}{2\sigma_2^2}\right)\right] \;\textnormal{.}\nonumber
\end{eqnarray}
Each fit has three free parameters: $\sigma_1, \sigma_2$ - the widths of each Gaussian and $p$ - the weight of the first one; the both means are zero. The parameters have been represented as functions of \mbox{$y = \log(r/r_M)$} with 2-degree polynomials $b·y^2+ c·y + d$; the values of the coefficients are given in Table 2. In Fig. 9 we compare the actual distributions of $\tau$ with the fitted functions for four distance intervals. The fits are satisfactory.

Next, we check whether the angular distribution $f_{\theta}(\vec\theta;r,E,s)$ obtained here for one $10^{17}$ eV iron shower is universal in the second sense, i.e. whether it does not fluctuate from one $10^{17}$ eV iron shower to another and whether it is independent of $E_0$ and the primary mass.
Fig. 10 illustrates that it is really the case. It shows the distributions $F_{\rho}$ and $F_{\tau}$ at two distances $r/r_M$ for four showers: initiated by proton or iron nucleus, each with  $E_0=10^{16}$ or $10^{17}$ eV, at $s= 0.95$.
It is seen that the curves are almost identical even for single showers with different primary parameters. 

For higher $E_0$  it is difficult to obtain $f (\vec\theta; r, E, s)$ from \emph{full} shower simulations - although the thinning procedure cuts the computing time, it introduces artificial fluctuations in the number of electrons in the  4-dimensional volumes $\Delta \theta_r \Delta \theta_t \Delta r\Delta E$, at the ages $s$, where the distributions are studied. However, we do not think it is actually necessary to get  $f(\vec\theta; r, E, s)$ for higher  $E_0$. Indeed, since we have shown that $f_r(r;E,s)$ is universal for $E_0=10^{16}\div 10^{20}$ eV in a broad range of $r$, $E$ and $s$ then also $f(\vec\theta; r, E, s)$ must be universal. If it was not true then it would be difficult to  imagine that  $f_r(r;E,s)$  would be universal, since the lateral and angular distributions depend on each other.

Concerning levels $s\ne 1$ it is only the functions $f (\vec\theta; r, E, s\ne 1)$ that need to be determined yet.
Since the angular distribution for a fixed $E$, \emph{integrated over the lateral distance} $r$, does not depend on the age $s$, and our model explaining the universality of $\vec\theta\cdot E^{0.73}$ holds for any $s$, we have reasons to postulate that at any $s$
\begin{equation}
   f_{\theta}(\vec\theta; r, E, s)\sim f'\left(\vec\theta\cdot E^{0.73}; \frac{r}{r'_{M}}\right) \;\textnormal{,}
\end{equation}
where 
\begin{equation}
r'_{M}=r_M\,e^{0.893(s-1)}\;\textnormal{.}
\end{equation}
This means that by rescaling the Moli\`ere radius according to Eq. (16) the distributions of $\rho$ and $\tau$ should stay almost the same. However, this has to be checked by simulations and will be done in another paper.

Finally, to understand the universality of the distribution of $\vec\theta\cdot E^{0.73}$ let us consider a simplified model of electron propagation. In the small-angle scattering approximation we have that
\begin{equation}
d\langle\theta^2_r\rangle =\frac{1}{2}\Big(\frac{21MeV}{E}\Big)^2\,dt\;\textnormal{,}
\end{equation} 
where $dt=dX/X_0$ is the element of the electron path length in radiation units. Assuming that the electron energy loss rate for bremsstrahlung and ionisation equals
\begin{equation}
-\frac{dE}{dt}=E+\varepsilon\;\textnormal{,}
\end{equation} 
where $\varepsilon$ is the critical energy of the air, and that the initial electron energy $E_i\gg E$, we obtain that
\begin{equation}
        \sqrt{\langle\theta^2_r(E)\rangle}\simeq\frac{21MeV}{\varepsilon}\sqrt{\frac{\varepsilon}{E}-\ln\,\left(1+\frac{\varepsilon}{E}\right)} \;\textnormal{.}
\end{equation}

This means that the variance $\langle\theta_r^2\rangle$ of an electron with energy $E$ depends on its final energy only.
In the energy region $0.2 \le E /\varepsilon \le 2$, where there are most electrons, the r. h. s. can be very well approximated by $\sim E^{-0.73}$ (Fig. 11). As it turns out from simulations it is not only $ \sqrt{\langle\theta_r^2(E)\rangle }\cdot E^{0.73}$ that is almost independent of $E$, but also the distributions of $\theta_r\cdot E^{0.73}$ (Fig. 6).

Unfortunately, it is not so easy to explain the universality of the distribution of $r\cdot E^{0.53}$. The electron lateral distance $r$, unlike its angle, does depend on the angles of the parent photons and electrons, so that the electron history is important. Thus, the lateral distribution of electrons with fixed $E$ does depend on the shower age.

\subsubsection{Approximate formula for the full electron distribution at shower maximum}

Summarising, at the considered level ($s \simeq$ 1) the number of electrons, $dN_e$, with energy \mbox{$(E, E + d\,E)$} at a lateral distance \mbox{$(r, r + d\,r)$} with angles \mbox{$(\theta_r , \theta_r + d\,\theta_r )$} and \mbox{$(\theta_t , \theta_t + d\,\theta_t )$} equals (Eq. 5) with our approximations
\begin{eqnarray}
dN_e&\simeq & N_e\,f_E\left(E;s\simeq 1\right)\,dE \cdot \nonumber \\
&& \cdot F_r\Big(\frac{r}{r'_E};s\simeq 1\Big)\,d\,\log\frac{r}{r'_E}\cdot \nonumber \\
&& \cdot F_{\rho}\Big(\theta_rE^{0.73}+3;\frac{r}{r_M},s\simeq 1\Big)E^{0.73} d\theta_r \cdot \nonumber\\
&&\cdot F_{\tau}\Big(\theta_tE^{0.73};\frac{r}{r_M},s\simeq 1\Big)E^{0.73}d\theta_t \;\textnormal{,}
\end{eqnarray}
where we have assumed that the distributions of $\theta_r$ and $\theta_t$ are independent.
\section{Shower reconstruction}
\subsection{The general idea of the method}
 We will show that the above described universality of extensive air showers enables an accurate prediction of their optical images registered in the consecutive time intervals while the shower develops in the atmosphere. The accuracy depends, of course, also on the knowledge of the atmospheric optical conditions and of the fate of the photons in the detector. The prediction becomes possible because the shapes of all the relevant electron distributions are the same for any high-energy shower, depending only on the shower age $s$. A high-energy shower means that the number of electrons at a given level $s$ in each 4-dimensional bin $\Delta E \cdot\Delta r\cdot \Delta \theta_r\cdot\Delta \theta_t$ is big enough as not to fluctuate much from shower to shower; as we have checked even $E_0\simeq 10^{16}$ eV fulfils this condition.

This work has been stimulated by our participation in the Pierre Auger Observatory (e.g. \cite{bib5,bib6}), so we shall assume that the optical detector is an imaging one, i.e. it consists of a mirror and a camera in the focal plane of it enabling one to measure the angular distribution of the arriving photons in a given time interval i.e. the instantaneous image of a shower. The telescope integration time is relatively short (100 ns in Auger) as compared to the total time while individual showers are in the telescope field of view. We assume that the shower geometry is known i.e. its distance from the telescope and the arriving direction. Thus, our data for one shower would consist of its consecutive optical images i.e. the angular distributions of photons at the telescope within successive 100 ns time intervals, as the air shower travels through the atmosphere.
The angular distribution of the photons arriving at the telescope translates, thanks to the concave mirror, to given numbers of photons in the particular pixels (PMTs) of the telescope camera.

Finding from air shower simulations the shapes of all relevant electron distributions on various levels one will be able, by assuming a particular absolute number of electrons $N_e(X(s))$, to predict the exact number of photons that should arrive at each camera pixel at any time interval. Allowing for the detector effects, connected with corrector ring, reflection efficiency of the mirror, PMT real field of view and its efficiency, and some others, one will be able to compare the predictions with the measurements and, by changing $ N_e(X(s))$, find the best fitting function. The method predicts equally well the Cherenkov and the fluorescence images so it is applicable to showers with any contribution of the former.

Our method can be applied to any individual air shower with $E_0 \gtrsim 10^{17} $eV emitting enough light to be detected from the side.

\subsection{The instantaneous optical image of an air shower}

\begin{figure}[t]
  \centering
  \includegraphics[width=0.45\textwidth]{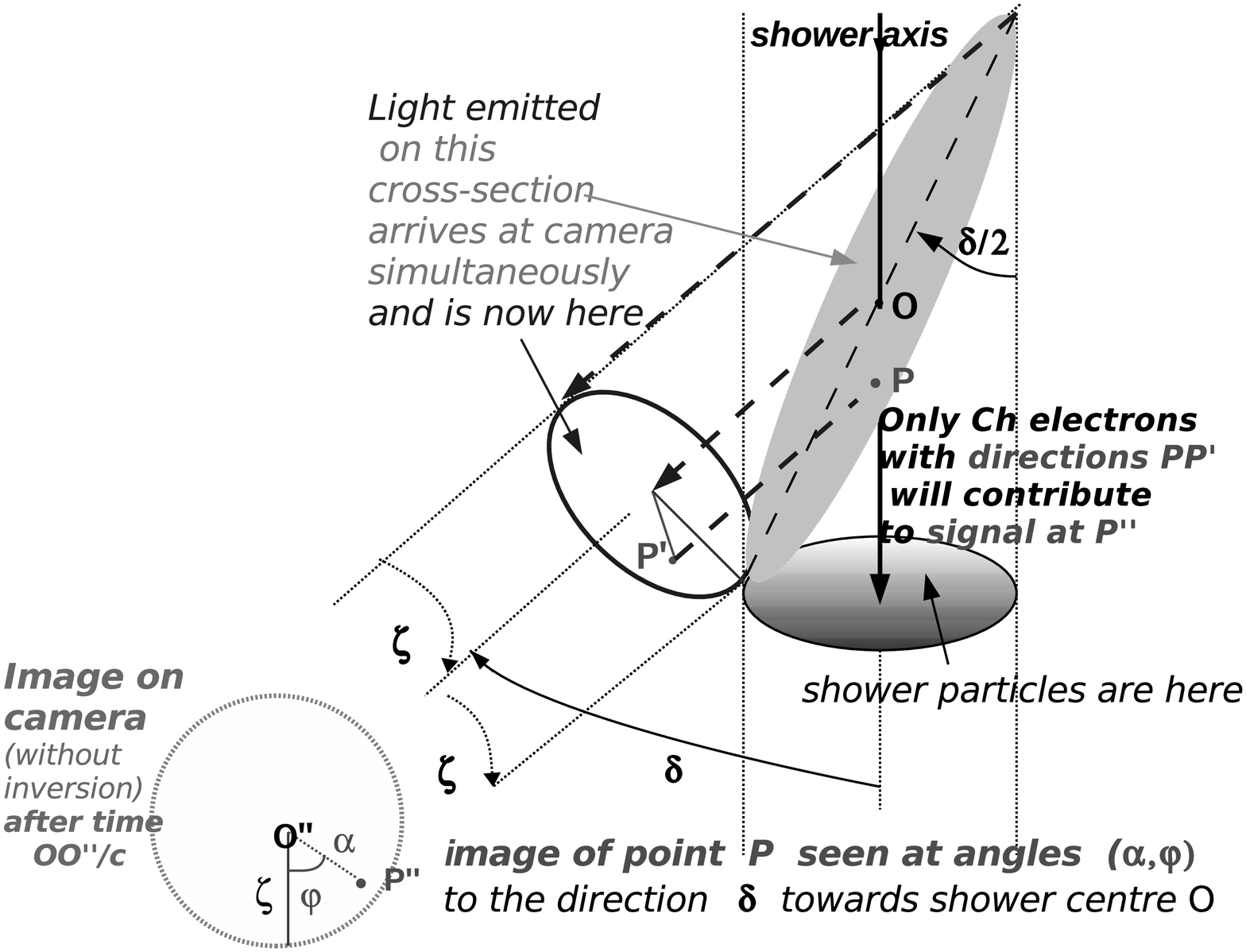}
  \caption{Schematic presentation of emission of light arriving simultaneously at camera.}
  \label{fig11}
\end{figure}

It was shown by Sommers \cite{bib23} that an instantaneous image of a shower has circular symmetry resulting from the axially symmetric lateral distribution of electrons. If the angle between the shower axis and the line of sight is $\delta$ then photons arriving simultaneously at the camera are those produced by particles at their crossing the plane inclined at an angle $\delta/2$ to the axis and perpendicular to the shower-detector plane (Fig. 12). As the lateral extension of the electrons is of the order of 1 Moli\`ere unit $\simeq 10\,g \,cm^{-2} $ the electrons at the upper end of the cross-section are almost at the same age $s$ as those at the lower end what results in the mentioned circular symmetry.

From Fig. 12 it is clear that we can assume that all electrons which give contribution to the shower image while the shower core is at the point $O$ are those lying at this level i.e. on the plane perpendicular to the shower axis at point $O$. The angular extent of the shower image in the light emitted at the observed level will depend, in general, on the lateral and angular distributions of the emitting electrons, and on the shower distance, of course. Each small surface element on the camera will receive photons emitted, or scattered, within a particular angle $d\vec\Omega$ by electrons at a particular distance element $r \,dr\, d\varphi$. To show the exact relations we define on the telescope camera two angles: $\alpha $ as the angular distance between the centre of the shower image and the point $P''$, where we want to predict the number of photons, and $\varphi$ the azimuthal position of point $P''$. It can be derived that photons arriving at point $P''$ must be produced at a lateral distance $r = \tan \,\alpha \cdot d$, where $d$ is the distance from the telescope to the shower, and azimuth $\varphi$ at angles $\theta_r$ and $\theta_t$ determined from the formulae
\begin{eqnarray}
\tan\,\theta_{r} &=& \tan\,\theta\cdot \cos\,(\varphi +\beta)   \\
\tan \,\theta_{t}&=& \tan\,\theta\cdot \sin\,(\varphi +\beta) \;\textnormal{;}  \nonumber
\end{eqnarray}
where
\begin{eqnarray}
\cos\,\theta&=&  \cos\,\delta\cdot \cos\,\alpha-\sin\,\delta\cdot \sin\,\alpha\cdot \cos\,\varphi  \\
\cos\,\beta&=& (\cos \,\alpha-\cos\, \delta\cdot \cos\,\theta)/(\sin\,\delta\cdot \sin\,\theta) \;\textnormal{.}  \nonumber
\end{eqnarray}
Thus, we see that it is not only the lateral and the angular distributions of electrons that have to be known to predict where exactly on the camera the produced photons fall. We would also need to know the angular distributions of electrons at various distances (see about the Cherenkov image below).

\subsubsection{The fluorescence image}
The fluorescence light is emitted according to the lateral distribution of the energy deposit, isotropically from each point of the air shower. Thus, the angular distribution of this light is determined by the lateral distribution of the energy deposit, projected on the plane containing point $P'$ in Fig. 12, perpendicular to the line of sight $OO''$, as seen from the telescope.

The lateral distribution of the energy deposited, as function of $r/r_M$, depends only on $s$. It was parametrized in \cite{bib24} and can be used for a prediction of the fluorescence image of an air shower once the shower age $s$ and the total number of electrons $N_e(X)$ (see section 3.2.4) at the considered level are known.

\subsubsection{The image in the scattered Cherenkov light}
The Cherenkov (Ch) photons travelling together with the shower particles are being scattered to the side in the atmosphere. Thus, the instantaneous shower image in this light depends on the lateral distribution of the Ch photons at the observed shower level and on their angular distribution after the scattering.

This problem was studied in \cite{bib18}, where it was shown that beyond the shower maximum the angular extent of the total optical image is dominated by Ch photons scattered at the observed level rather than by the fluorescence light. It was also shown that the lateral distribution of the Ch light scattered at the shower level $s$ depends mainly on the angular distributions of the emitting electrons above the scattering level. The electron lateral distributions there do not play an important role and were taken into account in an approximate way. It was assumed that the angular and lateral distributions of electron are independent of each other, which, as we have seen, is not true.

Strictly speaking, to accurately predict the lateral distribution of the Ch light arriving at some observed level $s$ it is necessary to know all the electron distributions discussed above, at the levels $s'$ higher that the observed one.
Indeed, the number of Ch photons , $dN_{Ch}$ \emph{produced} at the level $s'$ per unit path at a distance $(r, r+dr)$ at an angle $\vec\theta$ within $d\Omega$ equals
\begin{eqnarray}
&&dN_{Ch}(\vec\theta,r;s')=N_e\,dr\,d\Omega\cdot \\
&&\int_{E_{th(h)}}^{E_0}Y_{Ch}\,(E)f_E(E;s')\,f_r(r;E,s')\, f_{\theta}(\vec\theta;r,E,s')\,dE \;\textnormal{,}\nonumber
\end{eqnarray}
where $Y_{Ch}(E)$ is the number of Ch photons emitted by one electron with energy $E$ per unit path. Here we have assumed that the Ch light is produced in the direction of the electron. Approximate values of $dN_{Ch}$, assuming $\theta_r$ independent of $\theta_t$, are obtained with the help of Eq. (20).

Knowing the angular and lateral distributions of the \emph{produced} Ch photons at any level $s'<s$ and allowing for the atmospheric attenuation is sufficient to obtain the angular and lateral distributions of the Ch photons \emph{arriving} at the observation level $s$.
Finally, using formula (21) and (22) and knowing the angular distribution due to the scattering, one obtains, where at the camera the scattered Ch photons should arrive.

\subsubsection{The image in the direct Cherenkov light}
The direct Ch light is that \emph{produced} at the observed shower element $\Delta X$ towards the direction of the telescope. For most air showers detected by the Pierre Auger Observatory the contribution of this light to the optical image is very small. The elevation angles of the telescopes are not large : $0< \alpha<30^{\circ}$, which, together with the limit on the zenith angle $\theta_z < 60^{\circ}$ of the well reconstructed showers, results in rather large typical viewing angles $\delta$. Since the electron directions are concentrated around the shower axis and the Ch light is emitted at very small angles the direct light is small indeed.

However, the situation is different for the telescopes of the HEAT extension to the Auger Observatory \cite{bib25}, where the elevation angles reach $60 ^{\circ}$. We have derived the distribution of the viewing angles $\delta$ for a given elevation angle $\alpha$ assuming the isotropic zenith angle distribution of the shower directions, $F(\theta_z)$, cut above $60^{\circ} $. We have
\begin{equation}
   F(\theta_z)=\frac{2}{1-\cos^2\theta_0}\sin\,\theta_z\,\cos\,\theta_z \;\textnormal{,}
\end{equation}
where $\theta_0=60^{\circ}$. The resulting distribution $f(x)$, where $x=\cos\, \delta$ is the following

\subparagraph{a) $for \quad \alpha\le 30^{\circ}$}
\begin{eqnarray}
f(x)&=&  \frac{8}{3\pi}\Bigg\{ \sin\,\alpha\cdot x\, \left[\frac{\pi}{2}+\arcsin \left(\frac{\sin\,\alpha\cdot x-\frac{1}{2}}{\cos\,\alpha\sqrt{1-x^2}}\right)\right] \nonumber \\
 &&+ \sqrt{\cos^2\alpha-\frac{1}{4}+\sin\,\alpha\cdot x-x^2}\Bigg\} \;\textnormal{;}
\end{eqnarray}

\subparagraph{b) $for \quad\alpha\ge 30^{\circ}$}
\begin{eqnarray}
f(x)&=&  \frac{8}{3} \sin\,\alpha\cdot x \quad \textrm{for} \quad 0\le \delta\le \alpha-30^{\circ}      \\
f(x)&=& \textrm{as in case \textbf{a)} for }\alpha-30^{\circ}\le \delta\le 150^{\circ}-\alpha \;\textnormal{.}\nonumber
\end{eqnarray}
Fig. 13 shows the distribution $f(x)$ for $\alpha = 15^{\circ}$ - the elevation of the centre of an Auger telescope, $\alpha =45^{\circ}$ - that of the centre of a HEAT telescope and $\alpha =60^{\circ}$ - the HEAT maximum elevation. A dramatic difference between the $\delta$ distributions for Auger and the HEAT telescopes is seen for small angles $\delta$. For $\delta < 30^{\circ}$ the corresponding fractions of showers are $\sim $3.8$\%$, 21$\%$ and 29$\%$. (This is only a rough illustration of the differences since many selection criteria in data processing will surely influence these numbers). Nevertheless, Fig. 13 shows that the effect of the direct Ch light plays a considerably bigger role in the the air showers registered by HEAT and needs to be treated in an accurate way.

For distant air showers, when all Ch light falls into one pixel, the Ch signal is $\propto dN_{ch}(\delta )/d\Omega $, which is almost equal to the angular distribution of the electrons emitting Ch light integrated over the lateral distance $r$. However, if the shower is close enough so that its lateral extent can be measured by the telescope camera, the number of Ch photons registered by an individual pixel has to be calculated with the use of Eqs. (20)$\div$(22). 

 To illustrate the importance of this detailed approach we have calculated the value of the direct Ch signal produced at $s\sim 1 $ and observed from two distances such that the total image is contained within angle $\zeta$. Fig. 14 presents the ratio of the more accurately calculated Ch signal, i.e. using Eqs. (20)$\div$(22) to the approximate one $dN_{ch}(\delta )/d\Omega$. It is seen that even for $\zeta = 3.6^{\circ}$ the difference may be as big as $\sim 20\%$.

\begin{figure}[t]
  \centering
  \includegraphics[width=0.4\textwidth]{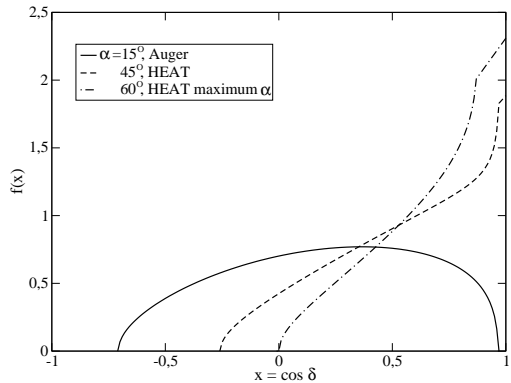}
  \caption{Distribution $f(\cos\delta)$ of shower viewing angle $\delta$, for three values of the elevation angle $\alpha$. Big difference is seen between an Auger telescope and that of HEAT.}
  \label{fig12}
\end{figure}

\begin{figure}[t]
  \centering
  \includegraphics[width=0.45\textwidth]{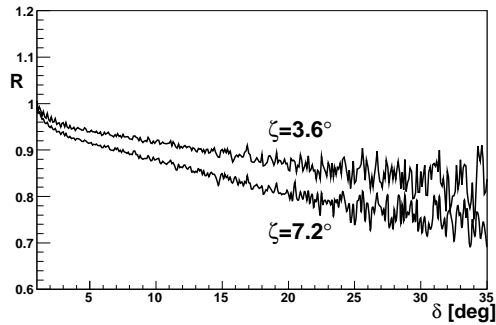}
  \caption{Ratio R of the accurately calculated direct Cherenkov signal to the approximate one, $d\,N_{Ch}(\delta)/d\,\Omega$, for two angular radii $\zeta$ of the Ch image as function of viewing angle $\delta$.}
  \label{fig13}
\end{figure}

\subsubsection{The total number of electrons}
So far we have discussed only the \emph{shapes} of the electron distributions. However, to predict the actual numbers of the photons arriving at the detector from a depth $X$ one needs to know the total number of electrons, $N_e(X)$, at this depth. It is well known that $N_e(X)$ can be described by a 4-parameter, $N_{max}, X_{max}, X_1$ and $\Lambda$, function of $X$, called the Gaiser-Hillas function \cite{bib26}. Thus, our final procedure of a shower reconstruction process is to find such values of the four parameters describing $N_e(X)$ with which the predicted numbers of photons in all hit pixels, at all time intervals, fit best the measured ones. The age $s$ of a level $X$ is determined solely by the ratio $X/X_{max}$ (Eq. 1).

\section{Conclusions}
   In this paper we have  extended the meaning of the universality of the electron distributions by showing that in the primary energy region $E_0=10^{16}\div 10^{17}$ eV the lateral distributions of electrons with any energy $E=0.02\div 1$ GeV on any level $s=0.7\div 1.3$ of the shower development can be described by one universal function $F(r/r_E,s)$, Eqs. (9)-(11).  We have called it the electron universality in the first sense.

We have also shown, by comparing our results with those of other authors, that the same function describes well the corresponding distributions in a broad range of the primary energy: $E_0=10^{16}\div 10^{20}$ eV, independently of the primary mass and thus, of the interaction model. This confirms the universality of the lateral distribution in the second sense.

Concerning the angular distributions $f(\vec\theta; r, E, s)$ we have found another universal (in the first sense) behaviour: at $s \simeq 1$ they depend on $\theta \cdot E^{0.73}$ only rather than on $\theta$ and $E$ separately. We have explained this by considering a model of the small angle scattering with simplified energy losses. We have checked that this is true for any shower with $E_0=10^{16}\div 10^{17}$ eV and we give arguments that this universality should extend up to $10^{20}$ eV.

Thus, we can state that  the function  $f (\vec\theta, r, E, s)$ describing fully the bulk of electrons in a high-energy shower is the same in any such shower independently of the primary energy and mass. The primary energy only has  to
be high enough, i.e. $E_0 \ge 10^{16}$ eV, so that the numbers of electrons in the 4-dimensional volumes $\Delta \theta_r \Delta \theta_t \Delta r\Delta E$, at the ages $s$, where $f (\vec\theta, r, E, s)$ is studied, be large. 

It should be, however, noted that the description of the electron distributions presented in this paper is not complete yet. There remains a better approximation of the angular distributions at particular distances $r$, since $θ\theta_r$ and $\theta_t$ are not entirely independent as we have assumed. Moreover, our hypothesis in Eq. (15) has to be checked for various shower ages $s$. This, however, does not undermine the universal character of the function $f (\vec\theta, r, E, s)$  in the second sense. It only means that  the minimum number of the independent variables describing all electron distributions in a shower has yet to be worked out.

The electron universality in the air showers can be used for reconstruction of longitudinal shower profiles from their optical images. The images can be predicted for any single shower, once the $N_e(X)$ is adopted because the shapes of the electron distributions do not fluctuate. 
A detailed knowledge of the function $f(\vec\theta, r, E, s)$, simplified here, is necessary for a correct prediction of the shower image in the Cherenkov light, which is of particular importance for air showers observed at viewing angles $\delta < 30^{\circ}$, as those registered by HEAT at Auger. For the fluorescence image it is only $f_r (r; E, s)$ that matters due to the isotropic emission of this light.

Our choice to determine electron distributions for fixed energies $E$ enables an easy allowing for the fact that air showers develop at various depths, thus at various heights $h$. Then, for the same shower age $s$ the energy threshold for the Cherenkov emission, $E_{th}(h)$, changes from shower to shower. To find the emitted number of Cherenkov photons from a given distance $r$, at some angle $\vec\theta$ one needs only to integrate the contributions from electrons with $E > E_{th}(h)$.

Another way of treating the Ch light would be to parametrise its emission as a function $F(\theta ; r, h, s)$. Nerling et al.\cite{bib21} chose this way by parametrising the distributions as $F( \theta; h, s)$, thus, integrated over the lateral distance $r$ and electron energy $E$. However, we think that the distributions $f(\vec\theta ; r, E, s)$ may be also of a more general interest than that connected with the Cherenkov light. It is by analysing them that we have found universal features of the electron distributions in a single air shower.

\paragraph{Acknowledgements}
The authors thank the Pierre Auger Collaboration for inspiring us to this study, teaching us the optical method and for many helpful discussions. We also thank Andrzej Kacperczyk for taking part in the determination of the lateral distributions for electrons with fixed energies.
This work has been supported by the grant no. N N202 200239 of the Polish National Science Centre.

\bibliographystyle{elsarticle-num}
\bibliography{<your-bib-database>}

\end{document}